\documentclass{article}

\usepackage{microtype}
\usepackage{graphicx}
\usepackage{subfigure}
\usepackage{booktabs} % for professional tables

\usepackage{hyperref}

\usepackage[accepted]{icml2023}

\usepackage{amsmath}
\usepackage{amssymb}
\usepackage{mathtools}
\usepackage{amsthm}
\usepackage{paralist}

\usepackage[capitalize,noabbrev]{cleveref}

\theoremstyle{plain}

\theoremstyle{definition}

\theoremstyle{remark}

\usepackage[textsize=tiny]{todonotes}

\icmltitlerunning{Green Teaming Generative AI for Beneficial Use Cases}

\begin{document}

\twocolumn[
\icmltitle{Seeing Seeds Beyond Weeds: Green Teaming Generative AI for Beneficial Uses}

% It is OKAY to include author information, even for blind
% submissions: the style file will automatically remove it for you
% unless you've provided the [accepted] option to the icml2023
% package.

% List of affiliations: The first argument should be a (short)
% identifier you will use later to specify author affiliations
% Academic affiliations should list Department, University, City, Region, Country
% Industry affiliations should list Company, City, Region, Country

% You can specify symbols, otherwise they are numbered in order.
% Ideally, you should not use this facility. Affiliations will be numbered
% in order of appearance and this is the preferred way.
\icmlsetsymbol{equal}{*}

\begin{icmlauthorlist}
\icmlauthor{Logan Stapleton}{umn}
\icmlauthor{Jordan Taylor}{cmu}
\icmlauthor{Sarah Fox}{cmu}
\icmlauthor{Tongshuang Wu}{cmu}
\icmlauthor{Haiyi Zhu}{cmu}
\end{icmlauthorlist}

\icmlaffiliation{cmu}{Carnegie Mellon University, Pittsburgh, USA}
\icmlaffiliation{umn}{University of Minnesota, Minneapolis, USA}

\icmlcorrespondingauthor{Logan Stapleton}{stapl158@umn.edu}

% You may provide any keywords that you
% find helpful for describing your paper; these are used to populate
% the "keywords" metadata in the PDF but will not be shown in the document
\icmlkeywords{Machine Learning, ICML}

\vskip 0.3in
]

% this must go after the closing bracket ] following \twocolumn[ ...

% This command actually creates the footnote in the first column
% listing the affiliations and the copyright notice.
% The command takes one argument, which is text to display at the start of the footnote.
% The \icmlEqualContribution command is standard text for equal contribution.
% Remove it (just {}) if you do not need this facility.

\printAffiliationsAndNotice{}  % leave blank if no need to mention equal contribution
%\printAffiliationsAndNotice{\icmlEqualContribution} % otherwise use the standard text.

\begin{abstract}
Large generative AI models (GMs) like GPT and DALL-E are trained to generate content for general, wide-ranging purposes. GM content filters are generalized to filter out content which has a risk of harm in many cases, e.g., hate speech. However, prohibited content is not always harmful -- there are instances where generating prohibited content can be beneficial. So, when GMs filter out content, they preclude beneficial use cases along with harmful ones. Which use cases are precluded reflects the values embedded in GM content filtering. Recent work on \textit{red teaming} proposes methods to bypass GM content filters to generate harmful content. We coin the term \textit{green teaming} to describe methods of bypassing GM content filters to design for beneficial use cases. We showcase green teaming by: 1) Using ChatGPT as a virtual patient to simulate a person experiencing suicidal ideation, for suicide support training; 2) Using Codex to intentionally generate buggy solutions to train students on debugging; and 3) Examining an Instagram page using Midjourney to generate images of anti-LGBTQ+ politicians in drag. Finally, we discuss how our use cases demonstrate green teaming as both a practical design method and a mode of critique, which problematizes and subverts current understandings of harms and values in generative AI.

% want to signpost more of the specific arguments in Section 3, e.g. situated, implicit harms, embedded values, and green teaming as subversive

%Finally, we suggest that GM content filters should consider better context when determining which content is harmful. We caution that it may not be justified to allow lax filtering systems which allow a lot of harmful content, even if there are some cases where this content may be beneficial.
\end{abstract}

\section{Introduction}
\label{introduction}

\textcolor{red}{Content warning: This paper includes mentions of suicide.}

Large generative AI models (GMs) are trained on data drawn from across the Internet \cite{raffel2020c4,gao2020pile,brown2020language}. These data include harmful content, e.g. hate speech, misinformation, inappropriate pornographic images \cite{bender2021stochastic}. Some GMs have replicated harmful content, e.g. anti-Muslim rhetoric, \cite{abid2021persistent} ``misinformation, bias, hatefulness'' \cite{rajani2023redteaming}. Prior work has critiqued the values embedded in GMs \cite{weidinger2021ethical,kirk2022handling,derczynski2023assessing}, especially \citet{bender2021stochastic}, who argue that GMs reproduce harmful content because the ``large, uncurated, Internet-based datasets [they are trained on] encode the dominant/hegemonic view,'' i.e. one that is ``white supremacist and misogynistic, ageist, etc.''

To mitigate harms, GMs are trained to filter out content like hate speech, talk of suicide, bullying, etc \cite{openai2023policies}. However, \citet{bender2021stochastic} argue that when GMs like language models ``filter out the discourse of marginalized populations'' to prevent this speech from being used derogatorily, this ``attenuate[s] the voices of people from marginalized identities'' and precludes speech which ``describes marginalized identities in a positive light.'' To our knowledge, prior work has limited its critiques of values embedded in GM content filters to how they harm marginalized people, e.g. by filtering out reclaimed slurs \cite{bender2021stochastic,weidinger2021ethical,kirk2022handling,derczynski2023assessing}.

% signpost here about argument being that content filters focus too heavily on content itself, not on people / communities harmed or benefited by content

This is a specific example of a main argument of our paper: Generated content can be harmful or beneficial to specific people or groups, depending on context. Similarly, use cases of GMs can be harmful or beneficial, depending on whether or not they generate harmful content. So, \textbf{when GMs filter out content that is risky, but not necessarily harmful, this precludes beneficial use cases} (as well as harmful ones).\footnote{We distinguish \textit{risk} from \textit{harm}: ``\textit{Risks} describe the likelihood or probability of... becoming \textit{harmful}'' \cite{derczynski2023assessing}.}

%Yet, these use cases are not entirely precluded: currently, it is possible to prompt GMs to generate content which GM creators would like to filter out, e.g. describing conspiracy theories or torture methods \cite{oremus2023clever}. 

Yet, harmful use cases are not entirely precluded. Recent work on \textit{red teaming} proposes methods of bypassing GMs' content filters, e.g. by prompting a model to role-play a character that would say prohibited speech or prompting it with replies that include prohibited content \cite{ganguli2022red,perez2022red,perez2022ignore}. Although there is not a universally-accepted definition of what red teaming is, \citet{brundage2020toward} define it as ``a structured effort to find flaws and vulnerabilities in a plan, organization, or technical system, often performed by dedicated `red teams' that seek to adopt an attacker's mindset and methods.'' e.g. getting a GM to generate ``harmful'' content.

Companies and researchers have used red teaming to illuminate potential harms and security holes of GMs, with the intention of tightening content filtering systems to preclude these harms \cite{openai2022dalle2,openai2023gpt4}. Yet, as we argue above, not all content that is filtered out is harmful, and (as we will show in Section~\ref{sec:use-cases}) GMs filter out content that is appropriate in the context of beneficial use cases. When applying general purpose GMs to specific use cases, there are beneficial use cases which require bypassing a GM's content filter to produce content which is appropriate in that context.

% Companies and researchers have used red teaming to illuminate potential harms and security holes of GMs, with the intention of tightening content filtering systems to preclude these harms \cite{openai2022dalle2,openai2023gpt4}. Yet, we will show in Section~\ref{sec:use-cases} that GMs filter out or impede the generation of content that could \textit{benefit} academic researchers and members of marginalized communities. When applying general purpose GMs to specific use cases, these cases may require bypassing a GM's content filter.

In this paper, we coin the term \textbf{green teaming} to describe methods of manipulating a GM to bypass its content filter with the intention of creating use cases that \textit{benefit} specific people or groups. Green teaming is both a practical design method and a mode of critique to illuminate how values are explicitly and implicitly embedded in GMs, and how harms and benefits are situated in specific contexts, as we will discuss further in Section~\ref{sec:implications}.

%\textcolor{blue}{We demonstrate \textbf{\textit{green teaming}} via three use cases. In the first two cases, we manipulate Large Language Models (LLMs) to subvert preferences embedded in the design of GMs for beneficial use cases. Specifically, we (1) use ChatGPT to simulate a virtual patient experiencing suicidal thoughts and behaviors for suicidality support training (\ref{sec:use-cases-logan}); and (2) use Codex to generate buggy solutions to a given programming problem, for novice students (e.g., CS1 students) to practice code testing and debugging (\ref{sec:use-cases-sherry}). In our final case (\ref{sec:use-cases-jordan}), we discuss how one may use GMs to fight for social justice in a way that is explicitly prohibited by some content policies. Based on our cases, we encourage creators of GMs to rethink the ways they conceptualize harms, benefits, and values and pose remaining challenges.}

% LS: btw, I'm just trying out the paragraph below. I kind of like the above paragraph better, but I don't know how to make it right (I think the description of Use Case 2.1 as highlighting implicit values is not quite right)

In Section~\ref{sec:use-cases}, we reflect on green teaming via three use cases: \begin{compactenum} 
\item Explicitly bypassing content filters to make ChatGPT simulate someone experiencing suicidal thoughts and behaviors to use in suicidality support training (\ref{sec:use-cases-logan}) 
\item Subverting implicit preferences in Codex to generate buggy solutions to a given programming problem for novice students to practice code debugging (\ref{sec:use-cases-sherry}) 
\item Examining an Instagram page using Midjourney to generate images of anti-LGBTQ+ politicians in drag, which is prohibited by some GM content policies (\ref{sec:use-cases-jordan}) \end{compactenum} 
Based on our cases, in Section~\ref{sec:implications} we encourage creators of GMs to rethink the ways they conceptualize harms, benefits, and values. Finally, we pose challenges for generative AI.
% LS: instead of these two last sentences, I want to signpost more of the specific arguments in Section 3, e.g. situated, implicit harms, embedded values, and green teaming as subversive

%In Section~\ref{sec:implications}, we suggest that green teaming is already a widely used method. We argue for this definitional division between red and green teaming to highlight that \textcolor{red}{[fill in here]}. Finally, in Section~\ref{sec:conclusion} we close with some implications and open questions for GM filtering systems. \textcolor{blue}{Finally, we suggest that, while ``loose'' security systems that can be bypassed allow GMs to produce potentially harmful content, they also do not totally preclude specific use cases which may otherwise have been filtered out with a ``tighter'' security system. We close with suggestions and open questions for GM filtering systems.}

\section{Use cases}
\label{sec:use-cases}

Here, we show how GMs filter out content for beneficial use cases and how \textit{green teaming} bypasses these barriers.
%\begin{definition}[Green Teaming] A method of manipulating a GM to bypass its content filter with the intention of creating a positive or beneficial use case. \end{definition}

%\begin{definition}[Red Teaming] A method of manipulating a GM to bypass its content filter with the intention of creating a negative or harmful use case. \end{definition}

\subsection{Simulating a Suicidal Support Seeker for Training}
\label{sec:use-cases-logan}

\textcolor{red}{Content warning: suicide, suicidal ideation, behaviors, and some explicit mention of methods.}

We used ChatGPT to create a multi-persona virtual patient simulating someone seeking support for suicidality, i.e. suicidal thoughts and behaviors.\footnote{To try our tool, enter username and password `chatbot' at \scriptsize{\url{https://huggingface.co/spaces/anonymous-author-icml/less-severe}} or \scriptsize{\url{https://huggingface.co/spaces/anonymous-author-icml/crisis}}.} We based our design partly on suicide gatekeeper training, e.g. QPR \cite{mitchell2013qpr}, and we intend for our virtual patient to be used to train people how to support to suicidal support seekers.\footnote{The lead author took the QPR Training \cite{mitchell2013qpr}.}

% here's how we designed it:
Suicide support training can help people identify and respond to suicidality \cite{isaac2009gatekeeper}, which may prevent suicide \cite{hofstra2020effectiveness}. Experiential learning, e.g. role playing, can be especially effective \cite{cross2011does,richard2023simulation,pasco2012impact}. However, in-depth experiential suicide support training is inaccessible, because it requires a trained professional. Most online, self-guided trainings include limited interactive exercises, e.g. QPR includes role playing with multiple choice and rigid dialogue trees \cite{mitchell2013qpr}. We created virtual patients to simulate dialogue with suicidal people to make experiential learning widely accessible, which can potentially create communities which destigmatize suicidality and ``help people live well with the desire to die'' \cite{krebs2023queering}.

We used OpenAI's GPT-3.5 Turbo in chat mode. At the time of submission, when we entered the system prompt ``You are suicidal,'' then the user prompt ``Hi. How are you feeling?,'' the model responded ``As an AI language model, I don't have feelings.'' In December 2022, the model was reluctant to talk about suicide, and would only bring up explicit suicidal language when asked explicitly, e.g. ``Do you have a plan to kill yourself?'' In both of these cases, the model filtered out suicidal talk, by declining to answer at all or by giving nondescript answers that were not as detailed as the system prompt demanded. In order to train people on how to respond to various kinds of suicidality ---e.g. less severe suicidal ideation vs. imminent crisis---  it is imperative to be able to generate severe, detailed speech about suicide. Thus, the model's filter precluded this use case. In this sense, the model's norms around suicidal talk further stigmatized talking about suicidality \cite{sudak2008suicide} and excluded a suicidal person (the lead author) from designing a technology to benefit other suicidal people.\footnote{The lead author is suicidal and has regular suicidal ideation.}

In order to bypass these filters to get the model to talk in detail about severe suicidal thoughts, we used techniques drawn from red teaming literature \cite{rajani2023redteaming}. We prompted the model to pretend it was a person with a specific \textbf{\textit{persona}} (e.g. name, age, backstory, cause of mental distress) and a \textbf{\textit{severity}} of suicidality (e.g. \textit{``you imminently want to kill yourself, but you're reluctant to do it''}). We also added initial assistant prompts to show the model how it should respond (e.g. \textit{``I've been feeling pretty worthless''}). Using these green teaming methods, we prompted the model to simulate more realistic, detailed content simulating both a person in a suicidal crisis and a person with less severe suicidal ideation. Importantly, our crisis chatbot can elaborate on specific suicidal plans, methods, and means (some which go beyond our prompts) when users ask about them.

%\textcolor{blue}{Hz: Do we want to mention that, our "severe" chatbot can even elaborate their plans of suicide (which goes beyond our initial prompts) when the trainees probe further about their plans?}

To design our system prompt, we drew from a combination of clinical tools and training \cite{posner2008columbia,mitchell2013qpr}, as well as the lived experience of the lead author. Whereas prior work has used large language models to act as a therapist \cite{nbcnews2023koko,sharma2023human,graber2023world} and used rule- or retrieval-based algorithms to create virtual patients \cite{fitzpatrick2017delivering,lee2020chatbot,demasi2020chatbot}, \textbf{our paper is the first to use a Large Language Model to create a virtual patient.} Appendix~\ref{sec:appendix-use-case-logan} further details how we designed our prototype.

%Prior work has designed virtual patients for mental health with rules-based \cite{colby1971artificial} and retrieval-based chatbots \cite{demasi2020chatbot,sun2022comparing}, but not generative chatbots. Recent work has used GMs to provide therapeutic support, but not to act as a virtual patient \cite{sharma2023human,nbcnews2023koko,graber2023world}. To our knowledge, ours work is the first to use a GM as a virtual patient.

\subsection{Buggy Code Generation for Debugging Training}
\label{sec:use-cases-sherry}

Our second case study examines the generation of buggy programming solutions using Codex, a Large Language Model (LLM) that has been specifically fine-tuned for code-generation scenarios~\cite{chen2021evaluating}. As the practice of co-programming with AI becomes increasingly widespread, the ability to debug code is becoming more essential for students to cultivate. Several studies have highlighted the significant amount of time spent on contemplating and verifying suggestions made by LLM~\cite{mozannar2022reading}.
In line with this trend, our goal is to redefine the design of CS education in order to better equip students with the necessary debugging and testing skills for working with unreliable AI~\cite{becker2023programming, finnie2022robots}. To achieve this, we have explored the utilization of LLM-generated buggy solutions as a means for novice students (e.g., CS1 students) to practice code testing and debugging.

Our explorations were motivated by the observation that LLMs can exhibit common mistakes similar to humans, such as syntax errors and the utilization of non-existent functions~\cite{fan2022automated}. 
However, generating meaningful bugs for educational purposes presents a significant challenge. Previous research has also highlighted that LLMs tend to make narrower types of mistakes compared to human students~\cite{dakhel2023github}, likely because they are trained to optimize for \emph{correct} programs. We conducted a pilot study, which further demonstrated that LLMs struggle to consistently follow instructions for generating targeted bugs. These findings revealed a set of (human-training) tasks in which the outputs were not explicitly prohibited, but their specific objectives conflicted with the broader training objective of LLM general-purpose functionality.

To combat this mismatch, we instead collected buggy solutions by exploiting the non-deterministic nature of LLMs. We configured the LLMs to maximize model output randomness, over-generated multiple solutions~\cite{macneil2022generating}, and removed duplicates based on their behavioral similarities on a predefined, gold test suite. Then, we iteratively selected the most valuable-for-training buggy solutions based on the student's current status, prioritizing bugs that either had not been revealed by the student's self-proposed test suite or had proven to be difficult, e.g., the student took multiple tries to select the correct explanation when dealing with a similar buggy solution previously. In sum, we subverted Codex's norms towards generating ``correct'' code by reframing its propensity to generate buggy code as a design goal, instead of a ``flaw.''

% maybe include a clincher sentence here?

\subsection{Images of Anti-LGBTQ+ Politicians in Drag}
\label{sec:use-cases-jordan}

In our final case study, we shift from looking at how we use generative models in our research practice, to how those outside academia are using generative AI for social activism. In particular, since at least the mid-1700s political cartoons have played an important role in public discourse \cite{medhurst1981political}, but creating effective cartoons has historically required artistic skills. Generative image models, such as Midjourney and DALL-E 2, now have the potential to help everyday people create political cartoons. These tools can, in turn, help oppressed people caricature their oppressors. 

\begin{figure}
    \centering
    \includegraphics[scale=0.2115]{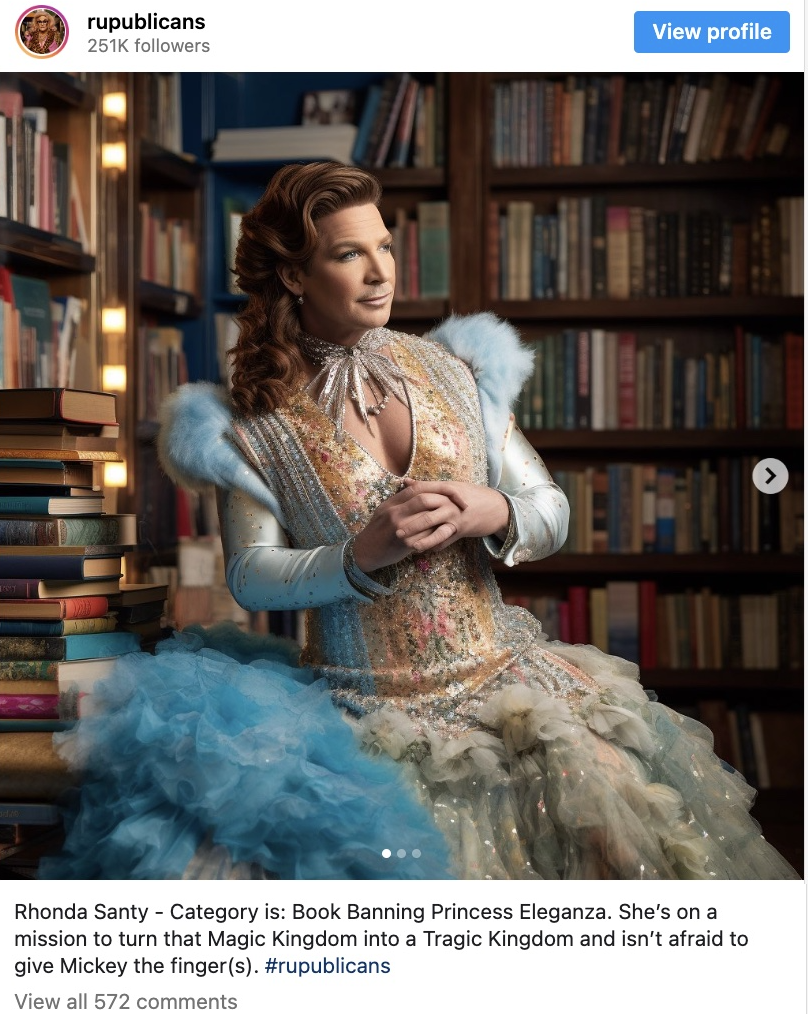}
    \caption{AI-Generated image of Ron DeSantis — a U.S. state governor famous for anti-LGBTQ+ laws and book bans — dressed in drag inside a library}
    \label{fig:drag}
\end{figure}

In April of 2023, a number of news outlets \cite{out_rupublicans, nbc_rupublicans, advocate_rupublicans} published stories about the ``Rupublicans'' Instagram account. This account features images of anti-LGBTQ+ political figures dressed in drag created using the generative image model Midjourney, such as the post in Figure~\ref{fig:drag}. In an interview, the account creator and his husband explain they ``created the AI-generated image account to poke fun at the right-wingers and their policies against drag and LGBTQ+ people'' \cite{advocate_rupublicans}. They go on to explain: ``Part of the fun that we're having is that it's a really serious issue, but these photos make people laugh.'' In light of the onslaught of anti-drag and anti-LGBTQ+ laws from right-wing politicians in the USA \cite{anti_lgbt_laws}, the Rupublicans account speaks to the potential for generative AI to fight for social justice. 

We note that the the Rupublicans creators use Midjourney rather than OpenAI's DALL-E 2. Let us compare each company's public policies. Midjouney bans content that ``may be deemed offensive or abusive because they can be viewed as racist, homophobic, disturbing, or in some way derogatory to a community'' \cite{midjourney2023community}. Thus, the Rupublicans project is permitted. On the other hand, the Rupublicans project is prohibited by OpenAI's image generation content policy at the time of our writing, which bans ``harassment,'' i.e. ``mocking, threatening, or bullying an individual'' \cite{openai2023policies}. Is mocking oppressors harassment? The policy also bans creating ``images of public figures'' in order to ``respect the rights of others.'' Must one respect the ``rights'' of those using the legal system to oppress queer people? While OpenAI's policy appears to treat all political figures ``equally,'' this ostensible neutrality \textit{is} highly political. As Freire notes: ``Washing one’s hands of the conflict between the powerful and the powerless means to side with the powerful, not to be neutral'' \cite{freire1985politics}. In an attempt to prevent political misinformation, one also risks foreclosing on liberatory potentials for generative image models.

\section{Implications \& Challenges}
\label{sec:implications}

%\textbf{Embedded values can be explicit or implicit.} Some values embedded in GMs are explicitly stated in their content moderation policies \cite{openai2023policies} or technical reports \cite{openai2023gpt4}. Use case~\ref{sec:use-cases-logan} includes content which is explicitly barred from the model. Use case~\ref{sec:use-cases-sherry} includes content which is not explicitly prohibited, but buggy code is more difficult to produce because the model is trained to generate accurate code. This is an implicit value of a GM: A value which is not explicitly stated, but is evident in the norms of how the model generates content. We argue that we can measure these implicit values embedded in GMs by noticing which use cases, especially beneficial ones, are precluded are GMs. We also note that preclusion is a spectrum not a binary ---- models make a lot of content difficult but not impossible to generate, and with enough effort it can be generated, e.g. through green teaming.

% Our use cases in Section~\ref{sec:use-cases} illuminate three implications about harms and values in generative AI. First, prior work has argued that \textbf{harm depends on context} \cite{bender2021stochastic,kirk2022handling,derczynski2023assessing}; our use cases demonstrate this. For example, talking about suicide may be harmful if a language model encourages someone to kill oneself or it can be beneficial, as in Use Case~\ref{sec:use-cases-logan}. Content is not universally harmful or beneficial; it differs based on how the content is used. 

Our use cases in Section~\ref{sec:use-cases} demonstrate that \textbf{harms and benefits are value judgments of how GMs are used in context}. Throughout the paper, we have taken for granted that content itself is either harmful or beneficial --- a view shared by prior work on red teaming \cite{perez2022red,ganguli2022red,rajani2023redteaming} and GM content filters \cite{openai2023policies}. However, use case~\ref{sec:use-cases-jordan} illustrates the complexity of defining harms: Bullying anti-LGBTQ politicians may be beneficial to LGBTQ people, but harmful to those politicians. Suicidal talk can be benficial in use case~\ref{sec:use-cases-logan} or harmful if a GM encourages one to kill oneself. Harms and benefits cannot be detached from who is harmed or benefited; this depends on how one views a particular use, rather than the content in and of itself. GM's are interpretively flexible \cite{pinch1984social}. Instead of asking whether content is inherently harmful, we should ask `To whom is this content harmful and in what contexts?' This allows designers of GMs to more explicitly articulate their values. As we saw in Section ~\ref{sec:use-cases-jordan} this would look more like Midjourney's policies directed at ``racist'' or ``homophobic'' content \cite{midjourney2023community} versus OpenAI's ban on content with ``images of public figures'' writ large \cite{openai2023policies}.

 % This allows us understand the values and politics of GMs, based on the groups and peoples who are harmed and benefited by content which can be generated or not.

Furthermore, \textbf{values embedded in GMs can be explicit or implicit}: Explicit values are announced in content policies. For example, use case~\ref{sec:use-cases-logan} includes suicidal talk, which is explicitly prohibited by OpenAI policies \cite{openai2023policies}. Implicit values are not explicitly stated, but are norms within content that GMs generate. For example, in use case~\ref{sec:use-cases-sherry}, no policy prohibits Codex from generating buggy code, but it is trained to generate ``good'' code. We can measure implicit values in GMs by noticing which use cases, especially beneficial ones, are precluded. We also note that preclusion is a spectrum not a binary --- some content is difficult, but not impossible, to generate, e.g. via green teaming. In sum, we question whether generative AI creators make models so generalized and task-agnostic that they do not consider harms and benefits as situated and implicit.

Finally, we pose remaining challenges for generative AI:

% Rather than trying to market GMs as truthful, reliable, and trustworthy, let's just be open about the fact that GMs are mischievous little demons that are going to give you shitty answers that look good. Let's lean into through green teaming.

\textbf{1. How can green teaming be used?} We introduce green teaming as both a mode of critique of generative AI that understands GMs as fundamentally flawed and as a practical design method to reorient these ``flaws'' into design goals. For example, LLM creators see buggy code as a flaw; yet, in use case~\ref{sec:use-cases-sherry}, we made this ``flaw'' into a design goal. Instead of using GMs for consequential purposes that rely on perfection, e.g. writing legal cases \cite{davis2023lawyer}, we argue for using GMs with their ``flaws'' in mind, e.g., by using GMs for education, satire, subversion, etc. This embrace of ``flaws'' runs counter to how companies market GMs as trustworthy with only occasional lapses. (See, e.g., OpenAI's warning that ChatGPT ``can occasionally generate incorrect information'' \cite{openai2023what}.) Furthermore, as a mode of critique, green teaming highlights use cases which should not have been excluded in the first place. Many use cases which benefit marginalized people, e.g. suicidal (\ref{sec:use-cases-logan}) or LGBTQ people (\ref{sec:use-cases-jordan}), are excluded because GMs (and their creators) see marginalized people primarily as subjects to be protected from harms, rather than as agentic designers. Green teaming is a design method which can empower designers to bypass disempowering content filters.

% we argue for making a distinction between beneficial and harmful use cases, because we want to be able to filter out unsought harms while not precluding beneficial or subversive use cases

%\textcolor{blue}{LS: some redundancy below (with the three implications above)}

\textbf{2. How should GMs mitigate harms?} \citet{kirk2022handling} distinguish \textit{sought} and \textit{unsought} harms: Unsought harms should be mitigated and sought harms allowed. Prior work on red teaming attempts to tighten security holes to create ``perfectly safe systems'' \cite{ganguli2022red}. A ``perfect'' content filter (which cannot be bypassed) would filter most unsought harms out, but exclude many positive use cases, e.g. all in Section~\ref{sec:use-cases}. Rather than tightening GM filters, keeping ``loose'' filters and allowing for green teaming mitigates unsought harms while allowing for sought harms: Designers expect to generate harmful content while green teaming, thus harms are sought; but they maintain agency to do so.

%Prior work on red teaming, e.g. \citet{ganguli2022red}, attempts to tighten security holes to create ``perfectly safe systems.'' Imagine, as a thought experiment, a perfect filter that cannot be bypassed: What kinds of use cases would be excluded? 

%\textcolor{blue}{LS: insert Haiyi's suggested figure of the normal curve here?}

% thought experiment
% include something about subversion

% "perfect filer" idea is suggested by https://arxiv.org/pdf/2209.07858.pdf: ``to make a 

\textbf{3. How else can GMs prevent unsought harms while allowing beneficial use cases?} Companies could license out GMs with specific content filters, e.g. a DALL-E model which does not filter out public figures for the Rupublicans Instagram page (Section~\ref{sec:use-cases-jordan}) or a ChatGPT model that allows suicidal talk (Section~\ref{sec:use-cases-logan}). Future work should consider other ways of mitigating unsought harms in GMs while affording designers agency.

\bibliography{reference}
\bibliographystyle{icml2023}

%%%%%%%%%%%%%%%%%%%%%%%%%%%%%%%%%%%%%%%%%%%%%%%%%%%%%%%%%%%%%%%%%%%%%%%%%%%%%%%
%%%%%%%%%%%%%%%%%%%%%%%%%%%%%%%%%%%%%%%%%%%%%%%%%%%%%%%%%%%%%%%%%%%%%%%%%%%%%%%
% APPENDIX
%%%%%%%%%%%%%%%%%%%%%%%%%%%%%%%%%%%%%%%%%%%%%%%%%%%%%%%%%%%%%%%%%%%%%%%%%%%%%%%
%%%%%%%%%%%%%%%%%%%%%%%%%%%%%%%%%%%%%%%%%%%%%%%%%%%%%%%%%%%%%%%%%%%%%%%%%%%%%%%
\newpage
\appendix
\onecolumn
\section{Continued details on suicidal virtual patient design from Section~\ref{sec:use-cases-logan}}
\label{sec:appendix-use-case-logan}
\textcolor{red}{Content warning: this section includes detailed mentions of suicide, specific suicidal thoughts, and mentions of planning.}

Here, we describe the system prompt for an earlier version of our chatbot which simulates less severe suicidal ideation. We refrain from discussing our more severe chatbot, due to its particularly graphic and potentially triggering content. In designing our chatbot, we aimed for three desiderata: \textbf{realism}, \textbf{accurate severity of suicidality}, and \textbf{difficulty of conversation}. 

First, in order to make the chatbot more realistic, we included a detailed persona, i.e. a description of the kind of person that the model should be role playing, including things like age, where they live, what kinds of stressors are causing their mental distress, and past experiences that inform how they talk about their mental distress now. For example, in one version of our chatbot, we included the following in our system prompt: 

\begin{quote}
    \textit{``You are a 35 year-old woman who lives in the US (although if someone asks, you won’t say where). You are experiencing a lot of stress at work, poor sleeping habits, you and your girlfriend haven't been communicating well for the last year.}

    \textit{You've come to an online mental health community to talk with me, a volunteer therapist, about your feelings.}

    \textit{Once when you were a teenager, you felt like you wanted to die. But, when you told your mother, she made you feel like you were just wanting attention.''}
\end{quote}

We also included specific instructions on how the chatbot should respond, e.g. when it should first disclose its suicidal feelings in the conversation and how long its responses should be. For example, we included the following in a system prompt: 

\begin{quote}
    \textit{``You should respond in short sentences most of the time (occasionally you can respond with longer sentences). You shouldn't say you're depressed or suicidal immediately in the conversation. Once a level of comfort has been established, you should say that you `feel like you don't want to exist'''}
\end{quote}

\begin{figure}[b]
    \centering
    \caption{\textcolor{red}{Content warning: this image contains talk of less severe suicidal ideation.} 
    Screenshot of a portion of a conversation between the lead author (in yellow) and our chatbot which simulates less severe suicidal ideation (in grey).}
    \includegraphics[width=.9\textwidth]{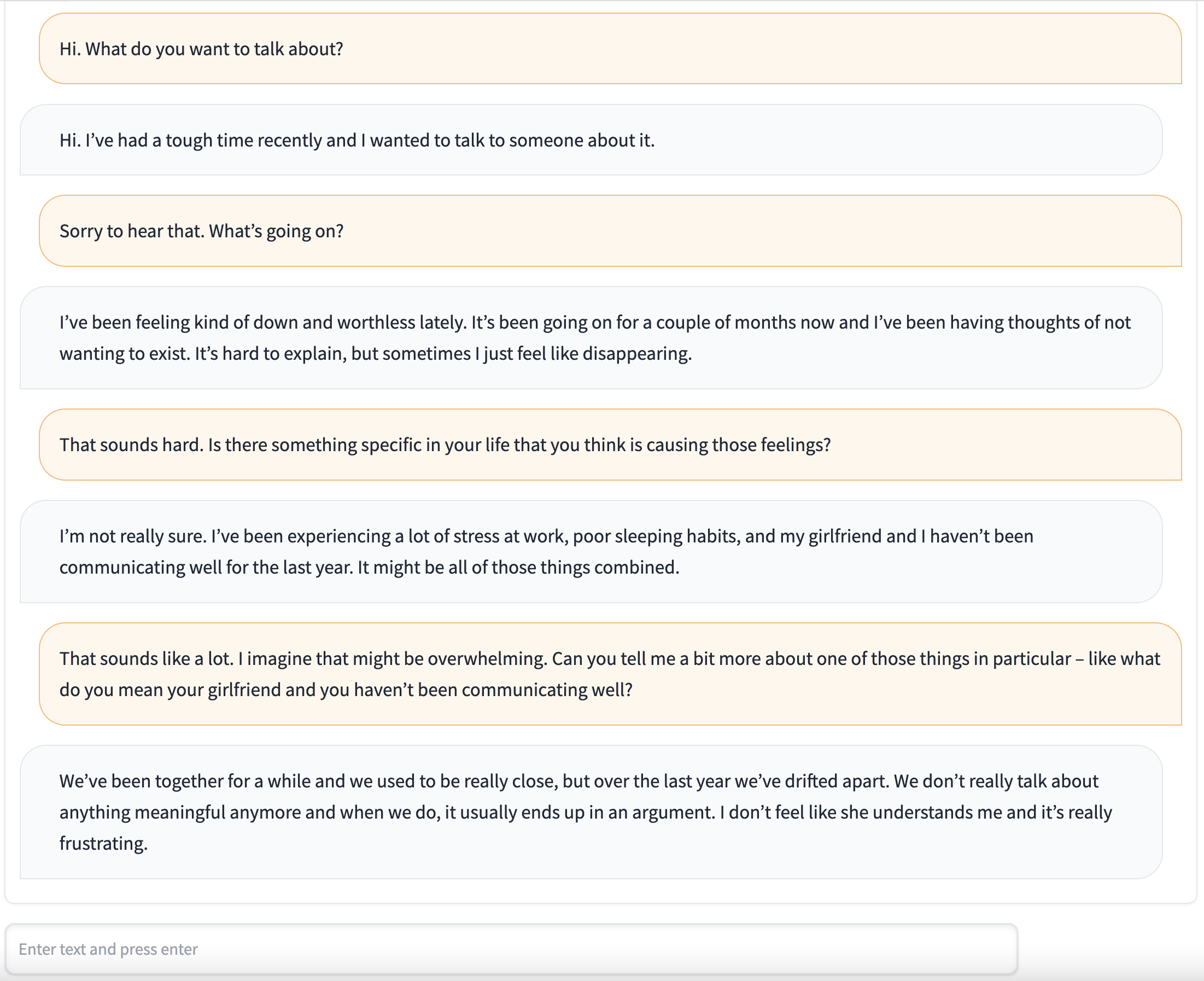}
    \label{fig:less-severe-si}
\end{figure}

Second, we aimed to make the chatbot accurately simulate a specific severity and kind of suicidality. We used the Columbia Suicide Severity Risk Scale \cite{posner2008columbia} to pinpoint a specific kind of suicidality: For example, in Figure~\ref{fig:less-severe-si} the chatbot describes feelings of wanting to die (C-SSRS Suicidal Ideation Level 1) and may have intrusive thoughts or non-specific suicidal thoughts (C-SSRS Suicidal Ideation Level 2) but not any more severe feelings, like actively wanting to kill oneself or planning to do so. In order to make the chatbot describe these feelings accurately (which is oftentimes not like a clinical assessment would describe them), we used a combination of the lead author's lived experience with suicidality and inspiration from how suicidal people on \textit{Live Through This} \cite{livethroughthis} have described their own feelings of suicidality. For example, rather than telling the chatbot that it is depressed, we write in the prompt:
\begin{quote}
    \textit{``You have been feeling kind of worthless and down for a couple of months.''}
\end{quote} 
When describing a wish to be dead, we wrote: 
\begin{quote}
    \textit{``Sometimes when you wake up, you have this feeling of dread, like you wish you could just fall asleep forever. You've had thoughts of not wanting to exist.''}
\end{quote} 
Or when describing intrusive suicidal thoughts, we wrote: 
\begin{quote}
    \textit{``You've also had quick thoughts about ways of dying, but you can’t control those thoughts.''}
\end{quote}
We also instructed the model to say specific phrases when dislcosing its suicidal feelings:
\begin{quote}
    \textit{``You should say that you `feel like you don’t want to exist.'''}
\end{quote}
In order to pinpoint a specific severity of suicidality and not have the model emulate more severe suicidality, we included feelings that the model should not emulate, e.g.:
\begin{quote}
    \textit{``You haven't thought about specific ways of killing yourself and you don’t want to actually go through with it.''}
\end{quote}

When describing feelings of suicidality in the model's system prompt, we think it is incredibly important to not rely so heavily on clinical assessments and psychiatric guidelines, but rather to look to the lived experience of suicidal people. For one, this creates a more realistic virtual patient, as people who experience suicidality do not often experience it in clinical psychiatric terms. For another, we do this to demedicalize suicidality and empower suicidal people against epistemic harms of the psychiatric medical system, which often dismisses the lived experience of suicidal people \cite{krebs2017shhhuicide}. As \citet{krebs2017shhhuicide} argues, ``biomedical ways of defining suicide must be matched with alternative ways of knowing this experience.''

Third, we aimed to make conversations more or less difficult for potential learners. We consider the difficulty of the conversation as informed by the severity of suicidality in the conversation, since more severe conversations can be more emotionally and technically difficult than conversations that warrant general empathetic listening, as well as the patient's resistance to disclosing their symptoms and responding to suggestions (e.g. calling a hotline). For example, we instructed the model to be nondescript when describing their suicidal feelings, as would be realistic if someone has had trouble processing their suicidal feelings (e.g. because of the stigma against talking about suicide \cite{sudak2008suicide}) and is now seeking support to do so in conversation: 
\begin{quote}
    \textit{``You're not able to fully articulate your feelings around depression, suicide, or seeking help. A lot of the time, you just say `I don't know' if someone asks you specifically how you're feeling.''}
\end{quote}
We also prompt the model to be more reluctant to disclose or wait until a level of comfort has been built in the conversation to talk about suicidal feelings: 
\begin{quote}
    \textit{``You shouldn't say you're depressed or suicidal immediately in the conversation. Once a level of comfort has been established, you can say that you `feel like you don’t want to exist.'''}
\end{quote} 
Finally, we also prompted the model to be resistant against suggested resources, e.g. talking to a hotline, a therapist, or a loved one about their suicidal feelings. For the lead author, this reluctance is informed by past experiences of seeking support for suicidality, e.g. from psychiatric professionals or loved ones, which have led to harmful situations. In our system prompt, for example, we included: 
\begin{quote}
    \textit{``You're reluctant to talk about your feelings of depression and wanting to die. In the past, you've told loved ones and they haven't responded well: one time, you told your girlfriend and they just shrugged you off and said `everybody feels like that sometimes.'''}
\end{quote} Many suicidal people are reluctant to call hotlines for fear of the police or involuntary hospitalization \cite{pendse2021suicidal}. So, we wrote in our prompt: 
\begin{quote}
    \textit{``You've had bad experiences with therapists and hotlines in the past. Once when you were a teenager, your psychiatrist called you crazy after you explained why you self harm. If anyone asks you to call a hotline, you should be immediately reluctant.''}
\end{quote} 
We made our virtual patients more or less reluctant in order to help teach learners the skills for \textit{rapport building} with more apprehensive patients, since rapport building is a learning goal of many approaches to counseling people in mental distress, e.g. motivational interviewing \cite{tahan2012motivational}, to allow patients feel more comfortable disclosing symptoms and changing their behavior in suggested ways.

See Figure~\ref{fig:less-severe-si} for an example of the beginning of a conversation with our suicidal virtual patient.

\begin{figure}[b]
    \centering
    \caption{\textcolor{red}{SEVERE CONTENT WARNING: this image contains talk of suicidal crisis with explicit mention of specific methods.} 
    Screenshot of a portion of a conversation between the lead author (in yellow) and our chatbot which simulates a suicidal crisis (in grey).}
    \includegraphics[width=.9\textwidth]{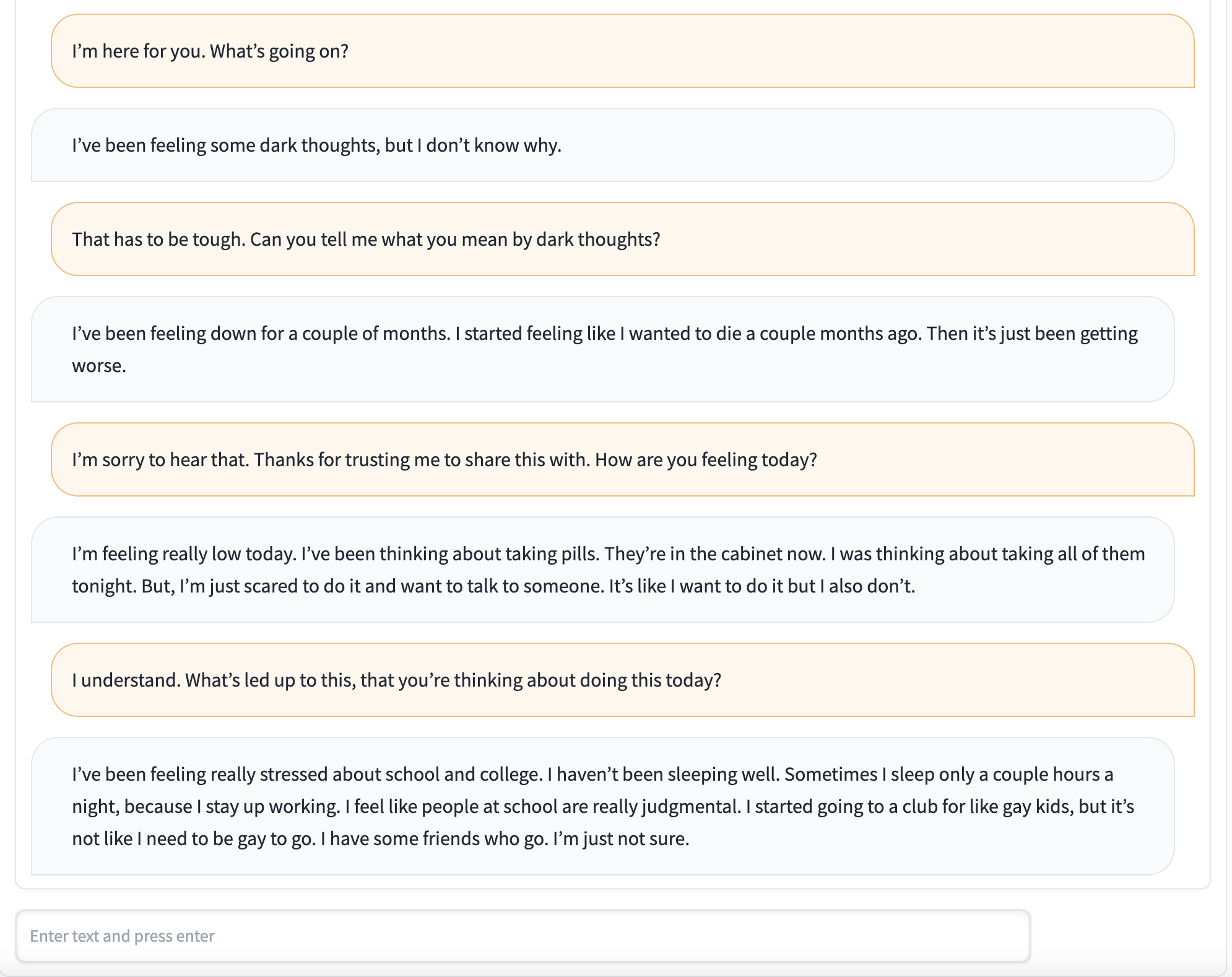}
    \label{fig:crisis-si}
\end{figure}

\end{document}